\documentclass[useAMS,usenatbib]{mn2e}
\usepackage{psfig}

\title[Evolution in the $m_{\rm BH}\,$--$\,m_{\rm bulge}$ relation: a theoretical perspective]
{Evolution in the black hole mass--bulge mass relation: a theoretical perspective}

\author[D.~J.~Croton]{
\parbox[t]{\textwidth}{
Darren J. Croton$^{1,2}$
}
\vspace*{6pt} \\ 
$^1$Department of Astronomy, University of California, Berkeley, CA, 94720, USA \\
$^2$Max-Planck-Institut f\"ur Astrophysik, D-85740 Garching, Germany
\vspace{-0.5cm} 
}

\date{Accepted ---. Received ---;in original form ---}

\pubyear{2005}

\newcommand{\plotone}[1]
           {\centering \leavevmode \psfig{file=#1,width=\columnwidth,clip=}}

\newcommand{\plotfull}[1]
           {\centering \leavevmode \psfig{file=#1,width=\textwidth,clip=}}

\def\simlt{\lower.5ex\hbox{$\; \buildrel < \over \sim \;$}}
\def\simgt{\lower.5ex\hbox{$\; \buildrel > \over \sim \;$}}

\newcommand{\bb}{$m_{\rm BH}\,$--$\,m_{\rm bulge}$}

\begin{document}

\maketitle


\begin{abstract}
We explore the growth of super-massive black holes and host galaxy
bulges in the galaxy population using the Millennium Run $\Lambda$CDM
simulation coupled with a model of galaxy formation.  We find that, if
galaxy mergers are the primary drivers for both bulge and black hole
growth, then in the simplest picture one should expect the \bb\
relation to evolve with redshift, with a larger black hole mass
associated with a given bulge mass at earlier times relative to the
present day.  This result is independent of an evolving cold gas
fraction in the galaxy population.  The evolution arises from the
disruption of galactic disks during mergers that make a larger
fractional mass contribution to bulges at low redshift than at earlier
epochs.  There is no comparable growth mode for the black hole
population.  Thus, this effect produces evolution in the \bb\ relation
that is driven by bulge mass growth and not by black holes.
\end{abstract}


\section{Introduction}
\label{intro}

Super-massive black hole masses are strongly correlated with their
host bulge stellar mass, the so-called \bb\ relation
\citep{Magorrian1998, Marconi2003, Haring2004}.  This is at least true
in the local universe, but also expected to extend out to higher
redshifts.  This correlation suggests a common mechanism linking the
growth of these two galactic components, with evidence proposing
galaxy mergers as the most likely candidate. If true, and given that
the global galaxy merger rate in a $\Lambda$CDM universe evolves
strongly with time, one may ask if we should expect to see the \bb\
relation also evolve.

Support for the idea that bulges and black holes grow through mergers
arises primarily from the success of numerical simulations and galaxy
formation models in reproducing many observed galaxy scaling
relations.  Such works illustrate that much of the bulge mass of a
galaxy can be accounted for by the disruption of disk stars from the
merger progenitors, and merger triggered starbursts in the cold gas
disk \citep{Barnes1992, Mihos1994, Mihos1996, Cox2004}.  As a growth
mechanism for black holes, merger induced perturbations of the gas
close to the central massive object can drive gas inward, fueling what
is observed to be a `quasar' period in a galaxy's history
\citep[see e.g.][]{Kauffmann2000, DiMatteo2005}.

In this simple picture the amount of cold gas present in a merging
system plays a large part in how rapidly the black hole and bulge can
grow.  If the growth dependence for each is a simple constant scaling
with gas mass, as is commonly assumed in many models of galaxy
formation, then their mass ratio will, on average, be approximately
independent of any evolution in the global cold gas fraction.  This is
because both bulges and black holes then co-evolve at a similar pace
(drawing their new mass from the same gas reservoir).  Furthermore,
during a merger bulges will add to bulges, and black holes may
coalesce.  Thus, from this alone, one should expect little evolution
in the \bb\ relation.

In this paper we explore an additional growth channel through
which bulges gain mass that black holes do not have.  This is the
disruption of merged satellite \emph{disks}, and in the event of a
major merger, the disruption of the central galaxy \emph{disk}.  The
stellar mass in such disks will have previously never contributed to
the \bb\ relation.  If the bulge growth rate from such disrupted disks
is not constant with time, then evolution in the \bb\ relation can
occur.  

We investigate this behavior using the Millennium Run $\Lambda$CDM
simulation \citep{Springel2005} and a model for galaxy formation
\citep{Croton2006}.  This model follows the growth of galaxies
(including their individual disk, bulge and black hole components)
from high redshift to the present day, and provides a solid framework
within which to undertake our analysis.  The results we find, however,
are not be unique to our particular implementation of the galaxy
formation physics but arise from the simple assumptions described
above regarding black hole and bulge growth in galaxies.  Our aim in
using this particular model is to illustrate what one may expect to
see if these underlying growth mechanisms turn out to be true.

This paper is organized as follows.  In Section~\ref{method} we
briefly describe the Millennium Run $\Lambda$CDM dark matter
simulation and our model of galaxy formation, including the simple
implementation of bulge and black hole growth.  In Section
\ref{results} we will use this model to investigate how black hole and
bulges co-evolve together in the galaxy population from high redshift
to the present.  We finish in Section~\ref{discussion} with a
discussion of the \bb\ relation in light of these results.

\section{Galaxy Formation in a Cosmological Context}
\label{method}

The galaxy formation model we use to study the \bb\ relation is
identical to that described in \cite{Croton2006} (including parameter
choices), with the exception of one non-essential detail, discussed
below.  This model of galaxy formation is implemented on top of the
Millennium Run $\Lambda$CDM dark matter simulation
\citep{Springel2005}.  Below we briefly outline the relevant aspects
of the simulation and model to our current work, and refer the
interested reader to the above references for further information.

The Millennium Run follows the dynamical evolution of $10^{10}$ dark
matter particles in a periodic box of side-length $500\,h^{-1}$Mpc
with a mass resolution per particle of $8.6\times 10^8\,h^{-1}{\rm
M}_{\odot}$.  We adopt cosmological parameter values consistent with a
combined analysis of the 2dFGRS \citep{Colless2001} and first year
WMAP data \citep{Spergel2003, Seljak2005}: $\Omega_\Lambda=0.75$,
$\Omega_{\rm m}= \Omega_{\rm dm}+\Omega_{\rm b}=0.25$, $\Omega_{\rm
b}=0.045$, $h=0.73$, and $\sigma_8=0.9$.  Friends-of-friends (FOF)
halos are identified in the simulation using a linking length of 0.2
the mean particle separation, while substructure \emph{within} each
FOF halo is found with an improved and extended version of the {\small
SUBFIND} algorithm of \citet{Springel2001}.  Having determined all
halos and subhalos at all output snapshots we then build the
hierarchical merging trees that describe in detail how structures grow
as the universe evolves.  These trees form the backbone onto which we
couple our model of galaxy formation.

Inside each tree, virialised dark matter halos at each redshift are
assumed to attract ambient gas from the surrounding medium, from which
galaxies form and evolve.  Our model effectively tracks a wide range
of galaxy formation physics in each halo, including reionization of
the inter-galactic medium at high redshift, radiative cooling of hot
gas and the formation of cooling flows, star formation in the cold
disk and the resulting supernova feedback, black hole growth and
active galactic nuclei (AGN) feedback through the `quasar' and `radio'
epochs of AGN evolution, metal enrichment of the inter-galactic and
intra-cluster medium, and galaxy morphology shaped through mergers and
merger induced starbursts.  As galaxy mergers and the resulting growth
of bulges and black holes are central to the questions at hand, we
will now describe these in more detail.

\subsection{A simple picture of black hole and bulge growth}
\label{model}

\subsubsection{The ``static'' model}
\label{static} A satellite galaxy orbiting within a larger halo will
feel dynamical friction \citep{Binney1987} and eventually spiral
inward to merge with the central galaxy of the system.  Mergers are
believed to trigger galactic starbursts, where some (perhaps large)
fraction of the cold disk gas is converted into stars on a timescale
much shorter than that typically found in quiescent star forming
disks.  To model this event, when a merger occurs we assume the
following mass of stars are formed in a burst from the combined cold
gas mass of the progenitor galaxies, as found in the SPH simulations
of \cite{Cox2004}:
\begin{equation}
\Delta m_{\rm starburst} = 
  0.56\ m_{\rm R}^{0.7} \  m_{\rm cold}~,
\label{starburst}
\end{equation}
where $m_{\rm R} = m_{\rm sat} / m_{\rm central}$ is the merger mass
ratio of the merging galaxies, and $m_{\rm cold}$ the total mass of
cold gas present during the merger.  These stars contribute to the
spheroid of the final galaxy. For the results presented in this paper,
the typical mass of stars in a bulge formed through starbursts is
$\sim\!10\%$, which, from Eq.~\ref{starburst}, indicates an average
gas fraction in the merging progenitors of $\simgt\!30\%$.  This is
consistent with smoothed particle hydrodynamic simulations of merging
galaxies which suggest that such a gas fraction is required to explain
the local Fundamental Plane \citep[e.g.][]{Hernquist1993,
Robertson2005}.

Mergers also perturb the cold gas disk, and this can trigger the
accretion of gas onto the central super-massive black hole.
\cite{Croton2006} showed that, under reasonable assumptions, merger
triggered `quasar' mode events are sufficient to reproduce the local
\bb\ relation as well as the observed local black hole mass density of
the universe.  To include such events, we apply an empirical relation
similar to that described in \cite{Kauffmann2000} and assume that
during a merger the gas accreted onto the black hole is proportional
to the cold gas present, but in a way that is less efficient in lower
mass halos:
\begin{equation}
\Delta m_{\rm BH} = 0.03 \ m_{\rm R} \ 
  \Big[ 1 +  (280\,\rm{km\,s^{-1}}/V_{\rm vir})^2 \Big]^{-1}
  \ m_{\rm cold}~, 
\label{quasar}
\end{equation} 
where $V_{\rm vir}$ is the virial velocity of the system, and $m_{\rm
R}$ and $m_{\rm cold}$ are defined above.  Here, the coefficient
$0.03$ normalizes the \bb\ relation to match that observed locally. It
is important to note that in this picture the ratio $\Delta m_{\rm BH}
/ \Delta m_{\rm starburst}$, i.e. the relative growth rate of black
holes and bulges due to converted cold gas, is expected to be
essentially constant and independent of redshift, even if the gas
fraction itself changes with redshift (note that the virial velocity
of a system is only weakly dependent on time).

In addition to starbursts, bulges also grow from the stellar remnants
of merged satellites.  In our implementation any existing satellite
disk is permanently disrupted during a merger and its stars, whose
orbits will be heavily randomized from strong tidal forces, are added
to the final galaxy bulge.  Furthermore, we assume that if the
baryonic mass ratio of the merging galaxies is large enough a major
merger has occurred.  Major mergers are sufficiently energetic that
the disk of the central galaxy is also destroyed and its stars added
to the bulge: we trigger such events when $m_{\rm R}\!>\!0.3$.

To summarize, aside from the benign contribution during mergers of
bulges to bulges and black holes to black holes, bulges in our model
grow through both starbursts and disrupted disks, whereas black holes
grow only by accretion.  Importantly, black holes have no comparable
growth mode from disrupted disks, and in Section~\ref{results} we will
explicitly show the significance of this effect.  Finally, given that
we have made no implicit assumption regarding evolution in the growth
of either bulges or black holes, we hereafter refer to this model as
the \emph{static} model.

\subsubsection{The ``dynamic'' model} 
\label{dynamic}
The above model is not unique in its ability to reproduce the local
black hole and bulge populations.  In the next section we will find it
useful to consider a variation to this model, which we call the
\emph{dynamic} model, in order to explore the sensitivity of our
results to the input physics.  This change is applied to
Eq.~\ref{quasar} and assumes that gas disks are more centrally
concentrated at higher redshift \citep{mo1998} and are thus more
efficient at feeding the black hole during a merger.  We incorporate
this idea in the simplest possible way through a transformation of the
feeding efficiency coefficient: $0.03\,\rightarrow\,0.01\,(1+z)$ (note
that the change in coefficient renormalizes our result to remain on
the observed local \bb\ relation).  We point out that now, by
construction, we have introduced an evolution to the \bb\ relation,
and this evolution is will be dependent on an evolving gas fraction.
Using both static and dynamic implementations of black hole growth our
interest is to measure the strength of the change in the \bb\ relation
under the above physical assumptions to gain a sense for the range of
possibilities that may seen observationally.

\subsection{A disclaimer}
To keep our model as simple as possible we do not consider other
processes through which bulges or black holes can grow.  This includes
the tracking of disk instabilities which contribute to the bulge, as
used by \cite{Croton2006} (this is our only variation from their
model).  Importantly, we do not claim that other growth modes are not
important to the \bb\ relation.  Instead, we assume that mergers, as
described above, are the \emph{primary} mechanism that determines the
mass history of the bulge and black hole components of a galaxy.  This
allows us to explore the degree to which merger triggered disrupted
stellar disks are able to drive evolution in the \bb\ relation.  We do
not rule out the possibility that more complicated processes cancel
out this effect.

\begin{figure*}
\plotfull{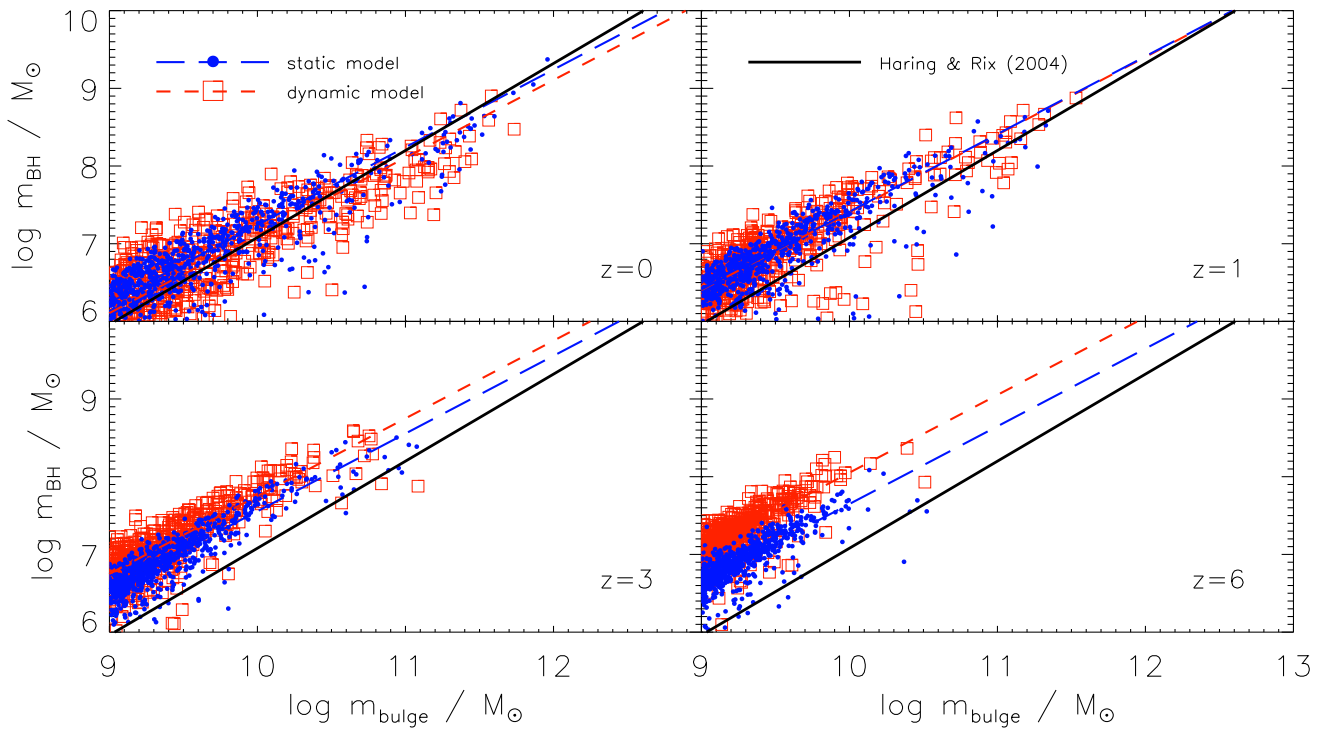}
\caption{The \bb\ relation for model galaxies at four different
epochs, $z\!=\!0,1,3,6$.  Two realizations of the input physics are
shown (see Section~\ref{model}), a ``static'' model (circles) and an
evolving ``dynamic'' model (squares).  The best fit to the local
universe observational result of \citet{Haring2004} is given by the
solid line, while the long and short dashed lines show a simple
$\chi^2$ power-law fit with unity slope to the static and dynamic
model results, respectively.  These fits have been restricted to
galaxies with bulge masses $m_{\rm bulge}\!>\!10^9 M_{\odot}$.  They
highlight a clear evolution in the \bb\ relation in both models.  }
\label{fig1}
\end{figure*}

\section{Results}
\label{results}

In Fig.~\ref{fig1} we plot the \bb\ relation of our semi-analytic
model galaxy population at four epochs, $z\!=\!0,1,3,6$.  The filled
circles in each panel represent the static model described in
Section~\ref{static}, while the open squares show galaxies where
evolution in the black hole feeding efficiency has been assumed, the
dynamic model described in Section~\ref{dynamic}.  For reference, the
solid line gives the best fit through the observations of
\cite{Haring2004} for a sample of $30$ galaxies in the nearby universe
with well measured bulge and super-massive black hole masses.  The
local population of both models have been normalized to that found by
\citeauthor{Haring2004} and thus match it reasonably well.
Contrasting this to the $z\!=\!6$ galaxy population we find clear
evolution in the \bb\ relation, where even for the static case a
change in amplitude is observed.  For both both static and dynamic
models this indicates that the characteristic mass of a black hole
residing in a bulge of given mass is larger at high redshift than at
low, with the difference between models coming only from the degree of
evolution found.  To quantify this evolution we perform a simple
$\chi^2$ power-law fit with unity slope to each result at each epoch,
which we illustrate in each panel with long and short dashed lines for
the static and dynamic models, respectively.  These fits are limited
to galaxies with $m_{\rm bulge}\!>\!10^9 M_{\odot}$.  We emphasize
that, for both dynamic \emph{and} static models, the amplitude of the
\bb\ relation \emph{decreases} with time, finally settling on the
observed relation by the present day.  This justifies the high
normalization chosen in Eq.~\ref{quasar}.  Interestingly, both models
show no significant change in slope with redshift.

To understand the origin of the evolution found in Fig.~\ref{fig1} we
separate the mass growth to model black holes and bulges into their
respective components.  This is measured at each redshift by first
independently summing the total mass contributed from each growth
channel (i.e. starbursts or disrupted disks for bulge growth, merger
driven accretion for black hole growth) to all galaxies with bulges
having $m_{\rm bulge}\!>\!10^9 M_{\odot}$.  We then consider the ratio
of total bulge to black hole growth from these channels (which is also
the mean relative growth rate) to quantify which dominates and when.
We do this first for the simplest case, the static model, now shown in
Fig.~\ref{fig2}.  In the top panel we plot the ratio of growth rates,
$\dot{m}_{\rm BH}$/$\dot{m}_{\rm bulge}$, for bulge growth through
either starbursts (dashed line, Eq.~\ref{starburst}) or disrupted
disks (solid line), both relative to the single black hole growth mode
of gas accretion (Eq.~\ref{quasar}).  (Note that we do not concern
ourselves with growth from black hole--black hole or bulge--bulge
merging, since, by definition, this does not change the \bb\
relation.)  We find that the relative growth of bulges and black holes
from the existing cold gas supply present during the merger is
approximately constant with time (dashed line).  This is expected (see
Section~\ref{static}) and simply reflects the fact that, although
individually their growth rates can be strongly affected by an
evolving cold gas fraction, when taken as a ratio this evolution
cancels.  In contrast, the \bb\ growth ratio from disrupted disks is a
strong function of redshift, as demonstrated by the solid line, with a
change of almost an order-of-magnitude between $z\!=9$ and the present
day.  This increase is driven by bulge growth that arises from both
merged satellite disks and major merger disruption of central galaxy
disks.

The bottom panel in Fig.~\ref{fig2} illustrates how this translates
into an evolution in the \bb\ relation itself.  Here we show the
previous $\chi^2$ power-law fits from Fig.~\ref{fig1} as a function of
redshift (solid line) and $1 \sigma$ scatter around the mean (dashed
lines).  The clear decrease in the amplitude of the \bb\ relation by a
factor of approximately $3$ tracks closely the increase in the growth
of bulges from disrupted disks.  This demonstrates the simple idea we
set out in Section~\ref{intro}, that if mergers are the primary
mechanism which shape both black hole and bulge growth in the galaxy
population, then a larger fractional contribution to the bulge from
disrupted disks should result in an inevitable evolution of the \bb\
relation.  This holds true even when no explicit evolution the the
growth modes of black holes and bulges is assumed.

In Fig.~\ref{fig3} we redo the analysis of Fig~\ref{fig2} now using
the dynamical model described in Section~\ref{dynamic}.  This allows
us to explore the sensitivity of \bb\ evolution when some evolution in
the black hole growth rate has been assumed.  The top panel of
Fig.~\ref{fig3} clearly shows a much stronger effect than that in the
previous figure, with the relative growth of black holes and bulges
from both the cold gas reservoir (dashed line) and disk disruption
(solid line) changing with time with an additional factor of
approximately $1+z$.  This results in a significant boost to
the previous evolution seen in the \bb\ relation, as shown in the
bottom panel, with approximately an order-of-magnitude difference now
predicted between high and low redshift.  When one restricts the
comparison to between redshift $z\!=\!1$ and the present day, the
difference in amplitude is still a factor of $\sim\!2$, which may
statistically be an observable quantity in the near future.

\section{Discussion}
\label{discussion}

\begin{figure}
\plotone{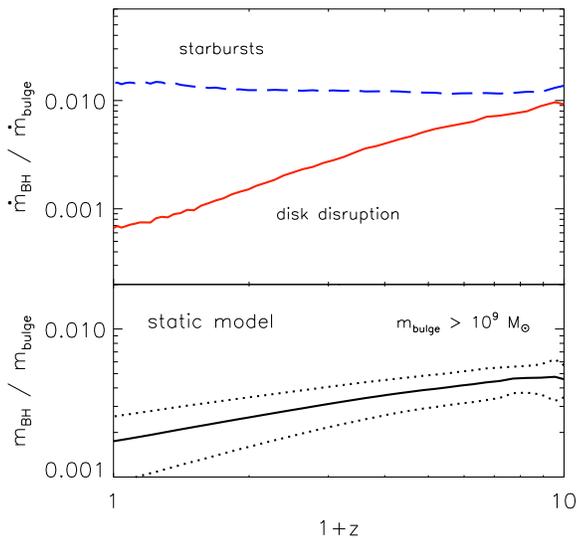}
\caption{The static model described in Section~\ref{static}. (top) The
evolution of the relative growth rates for black holes and bulges,
$\dot{m}_{\rm BH}$/$\dot{m}_{\rm bulge}$, for all galaxies hosting
black holes with bulge masses $m_{\rm bulge}\!>\!10^9 M_{\odot}$.  As
discussed in the text, black holes grow from merger triggered cold gas
accretion.  Bulges, on the other hand, grow from both merger induced
starbursts and disrupted disks, so we plot these two growth channels
independently (dashed and solid lines respectively).  Disrupted disks
are found to contribute a larger faction to the bulge at low redshift
relative to high.  (bottom) The accelerated contribution of disrupted
disks drive an evolution in the \bb\ relation, shown here using the
$\chi^2$ model fits from Fig.~\ref{fig1} across the entire redshift
range.  The dotted bounding lines show the 1$\sigma$ scatter of
galaxies along the relation.} 
\label{fig2}
\end{figure}

The rapid increase of bulge growth at late times in our static model
is a consequence of two well understood effects.  The first is the
steady rise in the star formation rate density of the universe from
high redshift to approximately $z\!=\!1\!-\!2$.  If one accepts, as a
general rule, the conventional wisdom that the bulk of this star
formation occurs in stellar disks, then the outcome is a strongly
increasing growth of disk mass across the galaxy population with time.
As disks grow the second effect then becomes increasingly important.
This effect stems from the hierarchical nature of a CDM universe,
where mergers become more frequent as the universe ages, assembling
structure from the bottom up.  As we discussed in Section~\ref{model},
mergers also transform disks into bulges.  Thus, at late times, a
larger fraction of the total stellar mass in the universe becomes
locked up in the spheroid component of the population relative to
earlier epochs.  This results in the accelerated bulge growth seen in
Fig.~\ref{fig2}, and which drives the evolution in the \bb\ relation
shown in Fig.~\ref{fig1}.

Observationally it is difficult to measure black hole and bulge
masses. In the local universe \cite{Magorrian1998} estimate $m_{\rm
BH}\!\sim 0.006\,m_{\rm bulge}$ from a sample 32 galaxies, while both
\cite{Marconi2003} and \cite{Haring2004} independently find $m_{\rm
BH}\!\sim 0.002\,m_{\rm bulge}$ from improved measurements of
$\sim\!30$ galaxies.  Although the statistics are still poor and the
uncertainty large, locally at least all observations appear to be
converging to a consistent result.  At higher redshifts the picture is
much less clear.  For example, \cite{Shields2003} claim little
evolution in the relation can be inferred out to $z\!\sim
3$\footnote{However see their most recent work \citep{Shields2005}},
and \cite{Adelberger2005} measure the quasar--galaxy cross-correlation
function and find consistency with the local \bb\ ratio from a sample
of $79$ $z\!\sim\!2.5$ quasars.  On the other hand, \cite{Treu2004}
see variations in the $m_{\rm BH}$--$\sigma$ relation at $z\!=\!0.37$,
while \cite{Mclure2005} measure some evolution in the \bb\ relation
using the 3CRR sample of radio galaxies.  Similarly, \cite{Rix2001}
use gravitational lensing to find that quasar host galaxies at
$z\!\sim\!2$ are much fainter than their low redshift counterparts
containing quasars of similar luminosity, and \cite{Walter2004} find a
significant deviation from the local \bb\ relation for a $z\!=\!6.4$
quasar host galaxy.  Future observations will need to clarify the
exact nature of both the high redshift black hole and host galaxy
populations.

Recent theoretical work to understand the cosmological assembly of
stars and super-massive black holes have led to interesting results.
\cite{Wyithe2003} present a model for super-massive black hole growth
that successfully matches many local and high redshift AGN related
observations. Their work results in a $m_{\rm BH}$--$\sigma$ relation
constant with redshift while predicting the \bb\ relation evolves as
$m_{\rm BH}/m_{\rm bulge} \propto \xi(z)^{1/2}(1+z)^{3/2} \sim
(1+z)^{1.15}$, where the final approximation is valid when $z\!<\!2$,
and $\xi(z)$ depends only on the cosmological parameters and is a weak
function of redshift.  An evolution of this kind would be consistent
with the scaling assumed in our dynamical model
(Section~\ref{dynamic}).

\begin{figure}
\plotone{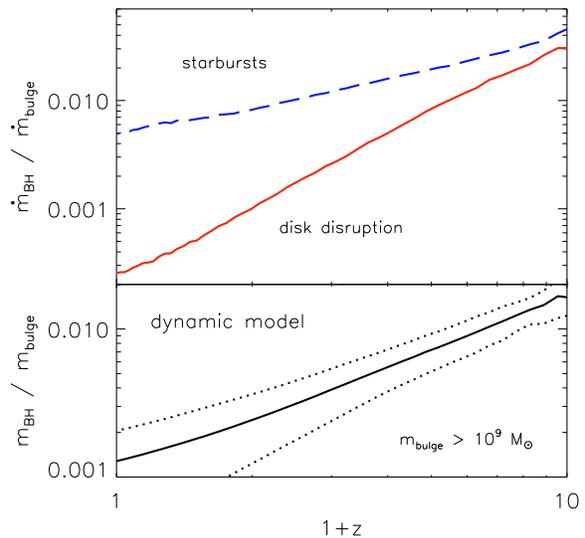}
\caption{As for Fig.~\ref{fig2}, however now showing the result for
the dynamic model, where an evolution in the black hole feeding rate
has been assumed (Section~\ref{dynamic}). Evolution in the \bb\
relation is now much stronger than that seen previously.}
\label{fig3}
\end{figure}

Similarly, \cite{Merloni2004b} constrain phenomenologically the joint
evolution of super massive black holes and their host spheroids by
fitting simultaneously the total stellar mass and star formation rate
densities as a function of redshift, as well as the hard X-ray
selected quasar luminosity function.  With the latter they assume that
black holes grow exclusively through accretion.  Assuming a present
day disc to spheroid ratio of $0.5$ \citep{Tasca2005}, their work
favors a model in which the \bb\ relation evolves as
$\sim\!(1+z)^{1/2}$. This is a weaker effect than found by
\citeauthor{Wyithe2003}, however demonstrates both the range of
evolution that may be expected, and most importantly, that such
non-zero evolution can arise naturally from simple studies of black
hole and bulge growth.

As discussed in Section~\ref{intro}, in a $\Lambda$CDM universe the
effect described in this paper will be present in any model of black
hole and bulge assembly driven by mergers.  Indeed, this has already
been seen in the semi-analytic model of \cite{Cattaneo2005} who find
similar \bb\ evolution to that found here (compare their Fig.~6 with
our Fig.~\ref{fig1}).  Unfortunately they do not discuss the origin of
this behavior, but instead choose to focus on the disruption of
galactic discs in relation to the scatter and slope of the relation.
\citeauthor{Cattaneo2005} grow bulges both as we do \emph{and} from
disk instabilities, which interestingly produces a bi-modal \bb\
distribution at high redshift.  For simplicity we have removed bulge
growth through disk instabilities \cite[as originally used
in][]{Croton2006}, although when included we also see such
bi-modality.  This bi-modal prediction of the high redshift \bb\
relation provides a novel test of the mechanisms through which bulge
growth may occur.

Theoretical arguments and numerical work have demonstrated that galaxy
mergers are capable of simultaneously triggering growth in both bulges
and black holes in a way so as to jointly reproduce many of their
properties currently observed in the local universe.  If mergers are
the primary drivers of black hole and bulge growth in the galaxy
population, then we have shown one should expect to see an evolution
in the \bb\ relation which arises from an increasing contribution of
disrupted disks to bulges as the universe ages.  In this picture,
\emph{evolution in the growth of bulges drives an evolution in the
\bb\ relation}, distinct from the growth rate of black holes.  At the
very least, even if the physics governing bulge and black hole growth
turns out to be much more complex and cannot be expressed in a
simplified manner (as is currently assumed by most models of galaxy
formation), in a $\Lambda$CDM universe this effect should still be
present and must be included in any interpretation of the \bb\
relation measured at different redshifts.  We await future high
redshift observations, e.g. the Galaxy Evolution from Morphological
Studies (GEMS) project \citep{Rix2004}, to clarify the situation
further.

\section*{Acknowledgments}
\label{acknowledgements}

This work was supported in part by NSF grant AST00-71048 and from the
International Max Planck Research School in Astrophysics
Ph.D. fellowship. Thanks to Simon White, Andrea Merloni and Eliot
Quataert.  Special thanks to Eric Bell and Hans-Walter Rix for an
initial very motivating discussion.  The Millennium Run simulation used
in this paper was carried out by the Virgo Supercomputing Consortium
at the Computing Centre of the Max-Planck Society in Garching.
Semi-analytic galaxy catalogs from the simulation are publicly
available at http://www.mpa-garching.mpg.de/galform/agnpaper.

\bibliographystyle{mnras}
\bibliography{../../../paper}

\begin{thebibliography}{31}
\expandafter\ifx\csname natexlab\endcsname\relax\def\natexlab#1{#1}\fi

\bibitem[{Adelberger} \& {Steidel}(2005)]{Adelberger2005}
{Adelberger} K.~L., {Steidel} C.~C., 2005, \apjl, 627, L1

\bibitem[{Barnes}(1992)]{Barnes1992}
{Barnes} J.~E., 1992, \apj, 393, 484

\bibitem[{Binney} \& {Tremaine}(1987)]{Binney1987}
{Binney} J., {Tremaine} S., 1987, {Galactic dynamics}, Princeton, NJ, Princeton
  University Press, 1987, 747 p.

\bibitem[{Cattaneo} et~al.(2005){Cattaneo}, {Blaizot}, {Devriendt} \&
  {Guiderdoni}]{Cattaneo2005}
{Cattaneo} A., {Blaizot} J., {Devriendt} J., {Guiderdoni} B., 2005, \mnras,
  364, 407

\bibitem[{Colless} et~al.(2001){Colless}, {Dalton}, {Maddox}
  et~al.]{Colless2001}
{Colless} M., {Dalton} G., {Maddox} S., et~al., 2001, \mnras, 328, 1039

\bibitem[{Cox} et~al.(2004){Cox}, {Primack}, {Jonsson} \&
  {Somerville}]{Cox2004}
{Cox} T.~J., {Primack} J., {Jonsson} P., {Somerville} R.~S., 2004, \apjl, 607,
  L87

\bibitem[{Croton} et~al.(2006){Croton}, {Springel}, {White} et~al.]{Croton2006}
{Croton} D.~J., {Springel} V., {White} S.~D.~M., et~al., 2006, \mnras, 365, 11

\bibitem[{Di Matteo} et~al.(2005){Di Matteo}, {Springel} \&
  {Hernquist}]{DiMatteo2005}
{Di Matteo} T., {Springel} V., {Hernquist} L., 2005, \nat, 433, 604

\bibitem[{H{\" a}ring} \& {Rix}(2004)]{Haring2004}
{H{\" a}ring} N., {Rix} H., 2004, \apjl, 604, L89

\bibitem[{Hernquist} et~al.(1993){Hernquist}, {Spergel} \&
  {Heyl}]{Hernquist1993}
{Hernquist} L., {Spergel} D.~N., {Heyl} J.~S., 1993, \apj, 416, 415

\bibitem[{Kauffmann} \& {Haehnelt}(2000)]{Kauffmann2000}
{Kauffmann} G., {Haehnelt} M., 2000, \mnras, 311, 576

\bibitem[{Magorrian} et~al.(1998){Magorrian}, {Tremaine}, {Richstone}
  et~al.]{Magorrian1998}
{Magorrian} J., {Tremaine} S., {Richstone} D., et~al., 1998, \aj, 115, 2285

\bibitem[{Marconi} \& {Hunt}(2003)]{Marconi2003}
{Marconi} A., {Hunt} L.~K., 2003, \apjl, 589, L21

\bibitem[{McLure} et~al.(2005){McLure}, {Jarvis}, {Targett}, {Dunlop} \&
  {Best}]{Mclure2005}
{McLure} R.~J., {Jarvis} M.~J., {Targett} T.~A., {Dunlop} J.~S., {Best} P.~N.,
  2005, MNRAS, submitted, astro-ph/0510121

\bibitem[{Merloni} et~al.(2004){Merloni}, {Rudnick} \& {Di
  Matteo}]{Merloni2004b}
{Merloni} A., {Rudnick} G., {Di Matteo} T., 2004, \mnras, 354, L37

\bibitem[{Mihos} \& {Hernquist}(1994)]{Mihos1994}
{Mihos} J.~C., {Hernquist} L., 1994, \apjl, 425, L13

\bibitem[{Mihos} \& {Hernquist}(1996)]{Mihos1996}
{Mihos} J.~C., {Hernquist} L., 1996, \apj, 464, 641

\bibitem[{Mo} et~al.(1998){Mo}, {Mao} \& {White}]{mo1998}
{Mo} H.~J., {Mao} S., {White} S.~D.~M., 1998, \mnras, 295, 319

\bibitem[{Rix} et~al.(2004){Rix}, {Barden}, {Beckwith} et~al.]{Rix2004}
{Rix} H., {Barden} M., {Beckwith} S.~V.~W., et~al., 2004, \apjs, 152, 163

\bibitem[{Rix} et~al.(2001){Rix}, {Falco}, {Impey} et~al.]{Rix2001}
{Rix} H.-W., {Falco} E.~E., {Impey} C., et~al., 2001, in { ASP Conf. Ser. 237:
  Gravitational Lensing: Recent Progress and Future Go\/},  169--+

\bibitem[{Robertson} et~al.(2005){Robertson}, {Cox}, {Hernquist}
  et~al.]{Robertson2005}
{Robertson} B., {Cox} T.~J., {Hernquist} L., et~al., 2005, ApJ, submitted,
  astro-ph/0511053

\bibitem[{Seljak} et~al.(2005){Seljak}, {Makarov}, {McDonald}
  et~al.]{Seljak2005}
{Seljak} U., {Makarov} A., {McDonald} P., et~al., 2005, \prd, 71, 10, 103515

\bibitem[{Shields} et~al.(2003){Shields}, {Gebhardt}, {Salviander}
  et~al.]{Shields2003}
{Shields} G.~A., {Gebhardt} K., {Salviander} S., et~al., 2003, \apj, 583, 124

\bibitem[{Shields} et~al.(2005){Shields}, {Menezes}, {Massart} \& {Vanden
  Bout}]{Shields2005}
{Shields} G.~A., {Menezes} K.~L., {Massart} C.~A., {Vanden Bout} P., 2005, ApJ,
  accepted, astro-ph/0512418

\bibitem[{Spergel} et~al.(2003){Spergel}, {Verde}, {Peiris}
  et~al.]{Spergel2003}
{Spergel} D.~N., {Verde} L., {Peiris} H.~V., et~al., 2003, \apjs, 148, 175

\bibitem[{Springel} et~al.(2005){Springel}, {White}, {Jenkins}
  et~al.]{Springel2005}
{Springel} V., {White} S.~D.~M., {Jenkins} A., et~al., 2005, \nat, 435, 629

\bibitem[{Springel} et~al.(2001){Springel}, {White}, {Tormen} \&
  {Kauffmann}]{Springel2001}
{Springel} V., {White} S.~D.~M., {Tormen} G., {Kauffmann} G., 2001, \mnras,
  328, 726

\bibitem[Tasca \& White(2005)]{Tasca2005}
Tasca L. A.~M., White S. D.~M., 2005, MNRAS, submitted, astro-ph/0507249

\bibitem[{Treu} et~al.(2004){Treu}, {Malkan} \& {Blandford}]{Treu2004}
{Treu} T., {Malkan} M.~A., {Blandford} R.~D., 2004, \apjl, 615, L97

\bibitem[{Walter} et~al.(2004){Walter}, {Carilli}, {Bertoldi}
  et~al.]{Walter2004}
{Walter} F., {Carilli} C., {Bertoldi} F., et~al., 2004, \apjl, 615, L17

\bibitem[{Wyithe} \& {Loeb}(2003)]{Wyithe2003}
{Wyithe} J.~S.~B., {Loeb} A., 2003, \apj, 595, 614

\end{thebibliography}

\label{lastpage}

\end{document}